\documentclass[genTeX]{nrc1}
\usepackage[pctex32]{graphicx}
\usepackage{epsfig}
\usepackage[figuresright]{rotating}
\usepackage{cite}
\usepackage{latexsym}
\def\tablefootnote#1{% 
\hbox to \textwidth{\hss\vbox{\hsize\captionwidth\footnotesize#1}\hss}}

\def\be{\begin{equation}}
\def\ee{\end{equation}}
\def\ba {\begin{eqnarray}}
\def\ea {\end{eqnarray}}
\def\nn {\nonumber}

\def\nn {\nonumber}
\def\a  {\alpha}
\def\b  {\beta}
\def\c  {\gamma}

\def\d  {\delta}
\def\D  {\Delta}

\def\l  {\lambda}

\def\n  {\nu}
\def\o  {\omega}
\def\O  {\Omega}
\def\p  {\pi}

\def\r  {\rho}

\def\la {\label}
\def\le {\left}
\def\ri {\right}

\def\f {\frac}

\def\bi {\begin{itemize}}
\def\ei {\end{itemize}}

\def\bc {\begin{center}}
\def\ec {\end{center}}

\journal{Can. J. Phys.}
\begin{document}

\title{Is entanglement entropy proportional to area?}
\author[Morteza Ahmadi]{Morteza Ahmadi}
\address{Department of Physics, University of Lethbridge, 
4401 University Drive, Lethbridge, Alberta T1K 3M4, Canada. 
\email{morteza.ahmadi@uleth.ca}}
\author[Saurya Das]{Saurya Das}
\address{Department of Physics, University of Lethbridge, 
4401 University Drive, Lethbridge, Alberta T1K 3M4, Canada. 
\email{saurya.das@uleth.ca}}
\author[S. Shankaranarayanan]{S. Shankaranarayanan}
\address{HEP Group, International Centre for Theoretical Physics,
Strada costiera 11, 34100 Trieste, Italy.
\email{shanki@ictp.trieste.it}}

\shortauthor{Ahmadi, Das and Shankaranarayanan}

\maketitle
\begin{abstract}
%Insert abstract here.
It is known that the entanglement entropy of a scalar field, found by
tracing over its degrees of freedom inside a sphere of radius 
${\cal R}$, is proportional to the area of the sphere (and not its
volume). This suggests that the origin of black hole entropy, also
proportional to its horizon area, may lie in the entanglement between
the degrees of freedom inside and outside the horizon. We examine
this proposal carefully by including excited states, 
to check probable deviations from the area law.
\\[0.8em]
\end{abstract}
\begin{resume}
Nous savons que l'entropie d'entrelacement d'un champ scalaire,
trouv\'ee en suivant ses degr\'es de libert\'e \`a l'int\'erieur
d'une sph\`ere de rayon ${\cal R}$, est proportionelle \`a
la surface de la sph\`ere (et non \`a son volume).
Ceci sugg\`ere que l'origine de l'entropie d'un trou noir,
\'egalement proportionelle \`a la surface de son horizon,
peut se trouver dans l'entrelacement entre les degr\'es de
libert\'e \`a l'int\'erior et \`a l'ext\'erior de l'horizon. 
Nous examinons avec soin cette 
hypoth\`ese, en incluant les \'etats excit\'es,
afin d'identifier les d\'eviations possibles de la loi de surface.

\traduit
\end{resume}

\section{Introduction}

There are strong indications that a black hole (BH) of mass $M$ and
horizon area $A_H$ possesses entropy and temperature, given
respectively by \cite{bh}:
\ba
S_{BH} = \f{A_H}{4\ell_{Pl}^2}\, , \qquad 
T_H = \f{\hbar c^3}{8\p GM} \, , \qquad \qquad  
\qquad ({\ell_{Pl} = \rm{Planck~length}}) \, .
\ea
The above entropy and temperature satisfy the laws of {\it BH
Thermodynamics}
\ba
T_H=\mbox{Constant on horizon}~,~~
d\le(Mc^2 \ri) = T_H dS_{BH} +  \mbox{work terms}~,~~ 
\D (S_{BH} + S_{matter} ) \geq 0 
\ea
where `work terms' are relevant for BHs with charge(s) and angular
momenta, and $S_{matter}$ refers to the entropy of matter outside the
BH. Usually the latter can be given a microscopic interpretation by
the relation $S_{matter} = \ln \O$, where $\O$ is the number of
accessible micro-states, compatible with the various macroscopic
parameters such as temperature, pressure and volume.  Although such an
interpretation for $S_{BH}$ is incomplete, important progress has been
made in various approaches to quantum gravity
\cite{sv,ash1,car1,adg1}. This, coupled with the area proportionality of 
BH entropy (as opposed to volume proportionality), has raised a 
fundamental question in quantum gravity,
namely: {\it What is the origin of BH entropy?} Potential candidates
include strings, $D$-branes, spin-network states and conformal
degrees of freedom at the horizon. Another popular (and also
incomplete) approach to the BH entropy is entanglement entropy, which
is a measure of the information loss due to the spatial separation
between the degrees of freedom inside and outside the horizon. The
so-called brick wall model has been a concrete realization of this
idea, in which entanglement entropy of the scalar (and other) fields
have been computed by tracing over their degrees of freedom outside
the horizon \cite{brickwall}\footnote{Note that tracing over the
outside does not pose any conceptual problem, since for a pure system,
tracing over a given subsystem and its complementary subsystem yield
identical entropies \cite{timmermans}.}. The brick-wall entropy turns
out to be proportional to the BH horizon area. The problem,
however, is that due to the infinite growth of the density of states
close to the horizon, one has to impose ultraviolet cut-off near the
horizon and hence the entropy depends on the cut-off scale. This
clearly is an undesirable feature.

A simpler physical system was considered in refs.\cite{BKLS,sred}, in
which the entropy of a scalar field on a suitably discretized {\it
flat} space was computed numerically, 
by tracing over the degrees of freedom of a hypothetical
sphere of radius ${\cal R}$. This gave the remarkable
result that the entanglement entropy was indeed proportional to the
area of the spherical surface (also see \cite{eisert} for 
analytical proofs of the result). 
This further supported the idea that
entanglement was responsible for BH entropy. One of the key
assumptions in refs.\cite{BKLS,sred} was that the harmonic oscillators
(HOs) resulting from the scalar field, on discretization of space, were
{\it all} in their ground states. In this article, we would like to
relax this assumption, and investigate the robustness of the
entropy-area relation in these physical systems. As a step towards
understanding of the entanglement entropy of $N$ coupled harmonic
oscillators (by tracing over $n < N$ degrees of freedom), in this work
we consider two HOs ($N=2, n=1$) in two physically interesting limits: 
coherent states and superposition of excited and ground 
states. We show, numerically, that the 
presence of excited states results in an increase of entropy. 
The generalization to arbitrary $N$ will be left to a future 
publication \cite{sdss}. 
 
In the next section, we will briefly review the entanglement entropy
for ground state HO. In section (\ref{secexcited}), we will generalize
the results for the two HO wave-function which is a superposition of
ground state and the first excited state, and show that the entropy
increases. In the concluding section, we will remark on the possible
implications of our results to the physically interesting case of a
large number of oscillators, whose couplings are determined by the
Lagrangian of a free scalar field. 

We will follow the notations of ref.\cite{sred} to provide easy
comparison, and henceforth use $\hbar=1$.

\section{Ground State Entanglement Entropy} 

The Hamiltonian for two coupled HOs of unit masses:
\be
 H = \f{1}{2} \le[ p_1^2 + p_2^2 + k_0 \le( x_1^2 
+ x_2^2 \ri) + k_1 \le( x_1 - x_2\ri)^2 \ri]  
\la{ham1}
\ee
has the ground state solution in terms of normal modes 
($x_\pm = \f{x_1\pm x_2}{\sqrt{2}},~
\o_+ = \sqrt{k_0}, \o_- = \sqrt{k_0+2k_1}$):
\be
\psi_0 \le( x_1, x_2\ri) 
=\psi_0 (x_+) \psi_0(x_-)
= \f{\le( \o_+\o_- \ri)^{1/4}}{\p^{1/2}}
\exp\le[- \le(\o_+ x_+^2 + \o_- x_-^2\ri)/2\ri] \, .
\ee
When traced over the oscillator characterized by $x_1$, the
resultant density matrix is:
\ba
 \r_{0} \le( x_2, x_2'\ri)
= \int_{-\infty}^\infty dx_1 \psi_0 \le( x_1, x_2\ri)
\psi_0^\star \le( x_1, x_2'\ri) 
= \sqrt{\f{\c-\b}{\p}}
\exp\le[ -\c\le(x_2^2+x_2'^2\ri)/2 +\b x_2 x_2'  \ri]~,
\la{dens1}
\ea
whose eigenfunctions and eigenvalues, respectively are
\ba
f_n(x) = H_n\le(\sqrt{\a} x \ri) \exp\le( - \a x^2/2\ri) \la{ef1} \, ,
\qquad \quad p_n = \le(1 -\xi\ri) \xi^n~ \, .
\la{ev1} 
\ea
The above density matrix gives rise to the following entropy \cite{BKLS,sred}: 
\be
S(\xi) = -Tr\le(\r_{0}\ln\r_{0}\ri) 
= -\sum_{n=0}^\infty p_n\ln p_n 
= -\ln(1-\xi) -\f{\xi}{1-\xi} \ln\xi \, , 
\ee
where 
\be
R^2 \equiv \f{\o_+}{\o_-} < 1,\quad
\a = \o_{-} R,\quad 
\b=\f{\o_-(1-R^2)^2}{4(1+R^2)}, \quad
\gamma = \f{1+6R^2+R^4}{4(1+R^2)},\quad
\xi = \le( \f{1-R}{1+R}\ri)^2 \, .
\ee
Note that $R = 0,1$ correspond to the strongly coupled and uncoupled
limits respectively.

The Hamiltonian for a free, massless scalar field $\varphi$ in flat
space-time is given by:
\be
H = \f{1}{2} \int d^3x \le[ \pi^2(\vec r) + |\nabla \varphi(\vec r)|^2\ri]~. 
\ee
Discretizing the space ($a$ being the lattice spacing, $Na$
signifying the infrared cutoff, and $l,m$ are the parameters in the spherical
harmonics $Y_{l,m}(\theta,\phi)$):
\be
H = \sum_{lm} H_{lm} = \f{1}{2a} \sum_{j=1}^N
\le[ \p_{lm,j}^2 + \le(j+\f{1}{2}\ri)^2
\le( \f{\varphi_{lm,j}}{j}
- \f{\varphi_{lm,j+1}}{j+1}
\ri)^2 
+\f{l(l+1)}{j^2} \varphi_{lm,j}^2 
\ri]~. 
\la{scham1}
\ee
The above expression for $H_{lm}$ 
is a special case of the general $N$-coupled
oscillator Hamiltonian:
\be
H = \f{1}{2} \sum_{i=1}^N p_i^2 
+\f{1}{2} \sum_{i,j=1}^N x_i K_{ij} x_j ~.
\ee
The corresponding $N$-HO ground state wave function
is given by:
\ba
\psi_0(x_1,\dots,x_n,x_{n+1},\dots,x_N) =
\le[\f{|\O|}{\p^N}\ri]^{1/4} 
\exp\le[ -\f{x^T\cdot \O \cdot x}{2} \ri] \, ,
\ea
where $\O^2 = K$. It can be shown that for the ground state wave
function, the density matrix (when one traces over $n<N$ oscillators)
can be factorized into a product of $(N-n)$ $2$-HO 
density matrices.  Thus
the total entropy is simply the sum of the entropies.  For the scalar
field, $n$ is taken to be proportional to the radius of the sphere
which is traced over, i. e., ${\cal R} = (n+1/2) a$.
For the Hamiltonian (\ref{scham1}) the interaction matrix
$K_{ij}$ can be read-off, resulting in the entanglement entropy 
\be
S = 0.3 (n+1/2)^2 \propto {\cal R}^2~,
\ee
signifying area proportionality.  It is worth noting, again, that
all the oscillators are assumed to be in their ground states.

\section{Entanglement entropy of excited states:} 
\la{secexcited}

Let us now consider the excited states of the $N$ HOs
discussed in the previous section. The corresponding wave-function is:
\ba
 \psi (x_1,\dots,x_n,x_{n+1},\dots,x_N) =
\le[\f{|\O|}{\p^N}\ri]^{1/4} 
~ \exp\le[ -\f{x^T\cdot \O \cdot x}{2} \ri] 
%\times
\prod_{i=1}^{N} 
\f{1}{\sqrt{2^{\nu_i} \nu_i!}}~
H_{\n_i} \le( {K_D^{\f{1}{4}}}_i~{\underbar x}_i\ri)~,
\ea
where $K_D \equiv U K U^T$ is a diagonal matrix ($U^TU=I_N$),
${\underbar x} \equiv Ux$ and $\nu_i \, (i=1 \dots N)$ are indices of
the Hermite polynomials.  The density matrix, tracing over first $n$
of $N$ oscillators, is
\ba
& & \!\!\!\!\!\!\!\!\!\!\!\!\!\!\!\!
\r_{0}\le( x_{n+1},\dots, x_N; x_{n+1}',\dots,x_N'\ri) 
= 
\le[\f{|\O|}{\p^N}\ri]^{1/2} 
\int \prod_{i=1}^n dx_i \exp\le[ -\f{x^T\cdot \O \cdot x}{2} \ri] 
\times \, 
\nn \\
%%%%%
&& 
%\int \prod_{i=1}^n dx_i \exp\le[ -\f{x^T\cdot \O \cdot x}{2} \ri] \, 
\prod_{i=1}^{N} 
\f{1}{\sqrt{2^{\nu_i} \nu_i!}}~
H_{\n_i} \le( {{K_{D_i}^{\f{1}{4}}}}~{\underbar x}_i\ri)  
{\exp\le[ -\f{x'^T\cdot \O \cdot x'}{2} \ri] \,
\prod_{j=1}^{N} 
\f{1}{\sqrt{2^{\nu_j} \nu_j!}}~
H_{\n_j} \le( {K_{D_j}}^{\f{1}{4}} ~{\underbar x}'_j\ri)}
\, .
\la{dens3}
\ea
The evaluation of the integral 
of the product of $2N$ Hermite polynomials, although 
may not be impossible, is in
general, non-trivial. In order to keep the calculations
simple, we consider two specific physical 
cases: (i) coherent states and (ii) superposition of ground and
first excited states.

The coherent states, which are eigenstates of the harmonic oscillator
annihilation operator with real eigenvalues, are described by the following
wave function:
\be
\psi_{CS} (x,a) \equiv \psi_0(x-a) = e^{{-i\hat p a}{}}~\psi_0 (x)~.
%= \sum_{n=0}^\infty a_n \psi_n(x).~~
\ee
The expectation of the position operator, w.r.t the coherent state wave
function, oscillates in time with an amplitude $a$ and the state has
the minimum allowable uncertainty
\be
\D p~\D x = \f{1}{2} \, , 
\ee
same as that of the ground state. For two coupled oscillators,
%described by the Hamiltonian (\ref{ham1}), one can choose the
the corresponding coherent state is:
\be
\psi_{CS}(x_1,x_2) \equiv  \psi_{CS} (x_+,a) \psi_{CS} (x_-,b)
= \psi_0( x_+-a) \psi_0 (x_--b) \, .
\ee
Defining
$
\tilde x_2 = x_2 - \le( {a-b}\ri)/{\sqrt{2}} \, ,
$
it is easy to show that the corresponding density matrix retains the 
same form as (\ref{dens1}), albeit in terms of these new variables:
\be
\r_{CS}(x_2,x_2') = \r_{out} \le( \tilde x_2,\tilde x_2'\ri) \, .
\ee
Thus, from Eqs.~(\ref{ev1}), it follows that the eigenfunctions are
$f_n({\tilde x})$ and eigenvalues remain unchanged ($p_n$), and we get
the interesting result that the entropy is the same as that for
the ground state! Presumably, this is because of the fact that 
coherent states are obtained by translating the ground state in phase
space. 
The result can be easily generalized to 
an arbitrary number of harmonic oscillators \cite{sdss}. 

Next, we consider the 
superposition of the ground and first excited 
state of the 2-HO system:  
\be
\psi(x_1,x_2) = \alpha_1~
\psi_1(x_+) \psi_0(x_-) + 
\beta_1~\psi_0(x_+) \psi_1(x_-)  + 
\gamma_1~\psi_0(x_+) \psi_0(x_-) \, 
 ~~[\a_1^2 + \b_1^2 + \gamma_1^2 = 1], 
\la{super1}
\ee
%
%where $\a_1^2 + \b_1^2 + \gamma_1^2 = 1$, 
%
\be
\rm{where}~~~\psi_n (x) = 
%\le( \f{\o}{\p} \ri)^{1/4} \f{1}{\sqrt{2^{n} n!}}
N_n (\o)~e^{ -\o^2 x^2/2} H_n(\sqrt{\o}~x),~
N_n (\o) =    
\le( \f{\o}{\p} \ri)^{1/4} \f{1}{\sqrt{2^{n} n!}}
\la{exc1}
\ee
is the $n^{th}$ excited state of an oscillator. Although from the
identity of particles one would expect $\a_1=\b_1$, we do not impose
such a condition at this point. From (\ref{dens3}), the density
matrix follows:
\ba
\r (x_2,x_2') =  \r_{0}(x_2,x_2') 
[ A \, (x_2^2+x_2'^2) + B \,  x_2 x_2' + C \, ( x_2 + x_2') + D ] \, ,
\la{dens2}
\ea
where $\r_0 (x_2,x_2')$ is the ground state density matrix given by
Eq.~(\ref{dens1}), and the constants are given as:
\ba
& & A = \a_1^2 a + \b_1^2 a_3 + \a_1\b_1 a_4 \, , \qquad  
B= \a_1^2b+\b_1^2b_3+ \a_1\b_1 b_4 \, , \qquad
C = \gamma_1 (\a_1a_6 + \b_1 a_7 ) \, , \nn \\
%%%%%
& & D= \a_1^2 c + \b_1^2 c_3 + \a_1\b_1 c_4 + \gamma^2 \, , \qquad   
a_6 =  \f{2\sqrt{\o_-}~R }{1+R^2} \, , \qquad  
a_7 = - \f{2\sqrt{\o_-}~R^2}{1+R^2} \, ,  \nn \\
%%%%%
\la{const1}
& & a = \f{R^2(1-R^2)(3+R^2)~\o_-}{4(1+R^2)^2} \, , \qquad  
b=\f{R^2(5+2R^2+R^4)~\o_-}{2(1+R^2)^2} \, ,\qquad  
c=\f{R^2}{1+R^2} \, , \\
%%%%%
& & a_3 = - \f{(1-R^2)(1+3R^2) \o_-}{4(1+R^2)^2} \, , \qquad
b_3 = \f{(1+2R^2+5R^4) \o_-}{2(1+R^2)^2} \, ,\qquad
c_3 = \f{1}{1+R^2} \, , \nn \\
%%%%
& & a_4 = \le(\f{1-R^2}{1+R^2} \ri)^2 \f{\o_- R}{2}\, ,\qquad
b_4 = -\f{R(1+6R^2+R^4) \o_-}{(1+R^2)^2}\, ,\qquad
c_4 = \f{2R}{1+R^2} \, . \nn  
\ea
It can be verified that:
$
Tr(\rho) =  \int_{-\infty}^\infty dx_2~\rho(x_2,x_2) 
= \a_1^2 + \b_1^2 + \gamma_1^2 = 1 \, .
$
To find the eigenvalues of the density matrix (\ref{dens2}), we follow
the general procedure outlined in ref. \cite{mf}. First, we expand
$\rho (x_2,x_2')$ in terms of general HO eigenstates (although,
in principle any complete set of functions should suffice):
\be
\r(x_2,x_2') = \sum_{n=0}^\infty h_n (x_2) g_n(x_2') 
~,~ h_m(x_2) = N_m (\a) 
\exp\le({-\f{\a x^2}{2}}\ri) H_m (\sqrt{\a}~x_2) ~.
\ee
Inverting, we get: 
{\small
\ba
g_m(x_2')&=& \int_{-\infty}^\infty dx_2 \, \r(x_2,x_2') \, h_m (x_2) \\
&=& p_m N_m e^{-\f{\gamma x_2'^2}{2} \f{(\b x_2')^2}{2(\gamma+\a)} } 
~\le[ \le(B_1 x_2'+E_1 \ri)  H_{m+1} (\sqrt{\a}~ x_2') + 
\le( C_1 x_2'^2 + D_1 + F_1x_2' \ri) H_m (\sqrt{\a}~x_2')  \ri] ~, \nn
\ea
}
where
{\small
\ba
& & \!\!\!\!\!\!\!\!\!\!\!\!
B_1 = -\sqrt{\a} \le[ \f{2{\bar a} \gamma }{\gamma^2-\a^2} 
+ \f{\bar b}{\sqrt{\gamma^2-\a^2}} \ri] \, ,
\qquad C_1 = \f{2{\bar a} \gamma}{\gamma-\a} 
+ {\bar b} \sqrt{\f{\gamma+\a}{\gamma-a}} \, ,
\qquad D_1 \equiv D_{11} + D_{12}, \nn \\
%%%%%
& & \!\!\!\!\!\!\!\!\!\!\!\!
D_{11} = \f{\bar a}{\gamma+\a} + {\bar c} \, ,
\qquad D_{12}  = - \f{2{\bar a} \a}{\gamma^2-\a^2} \, ,
\qquad E= - \f{{\bar d} \sqrt{\a}}{\sqrt{\gamma^2-\a^2}} \, ,
\qquad F_1 = {\bar d} \le[ 1+ \sqrt{\f{\gamma+\a}{\gamma-\a} }\ri]\\
%%%%
& & \!\!\!\!\!\!\!\!\!\!\!\!
{\bar a}= \a_1^2 a +  \b_1^2 a_3 + \a_1\b_1 a_4,~
{\bar b} = \a_1^2 b + \b_1^2 b_3 + \a_1 \b_1 b_4,
~{\bar c} = \a_1^2 c + \b_1^2 c_3 + \a_1 \b_1 c_4 + \gamma^2,
~ {\bar d} = \a_1 a_6 + \b_1 a_7~,
\nn
\ea
}
\noindent and $a_i, b_i$ have been defined in Eq.~(\ref{const1}). 
The next step is to define the matrix equivalent of $\r$, i. e.,
{\small
\ba
\a_{pm} &\equiv & \int_{-\infty}^\infty dx~g_m(x) h_p(x) \\
%%%%
&=& p_m \le[
\le( D_{11} + p D_{12} + \f{B_1(p+1)}{\sqrt{\a}} + 
\f{C_1(2p+1)}{2\a} \ri) \d_{pm} + \f{C_1}{2\a} \sqrt{(p+1)(p+2)} 
\d_{p,m-2} \right. \nn \\
%%%%%
&+& \left.  \sqrt{p(p-1)}\le( \f{B_1}{\sqrt{\a}} + \f{C_1}{2\a} \ri)
\d_{p,m+2} + F_1 \sqrt{\f{p+1}{2\a} }\d_{p+1,m} 
+ \le(E_1 \sqrt{2p} + F_1 \sqrt{\f{p}{2\a}} 
\d_{p-1,m}\ri) \d_{p-1,m}\ri] \, .\nn
\ea
}
Although formally diagonalizable, the eigenvalues $\l_p$ of 
the above penta-diagonal matrix 
are most easily found numerically. With MAPLE, using 
upto $40 \times 40$ matrices, we verified that it has unit trace.
\be
Tr(\a_{pm}) = \sum_{m=0}^\infty \a_{mm} =1\, , \qquad
\a_{mm} = p_m \le[ \le( D_1 + \f{B_1}{\sqrt{\a}} + \f{C_1}{2\a} \ri)
+ m \le( D_{12} + \f{B_1}{\sqrt{\a}}+ \f{C_1}{\a} \ri) \ri] ~. 
\ee
%
%where $\l_p$ are the eigenvalues of $\a_{pm}$ matrix. 
The corresponding entropy as function of $\a_1,\b_1,R$
defined as:
\be
S(\a_1,\b_1,R) = -\sum_{p=0}^\infty \l_p \ln \l_p 
\ee
was also computed numerically, and for all $\a_1,\b_1 \neq 0$ it was
found that $S(\a_1,\b_1, R) \geq S(0,0,R)$, where $S(0,0,R)$ is the
ground state entropy. The equality holds {\it only} in the uncoupled
limit $R=1$ and $\a_1=\b_1$. 
These features are visible in Fig.(1), where 
we have plotted entropies for the excited state
[$\a_1=\b_1=1/\sqrt{2},\gamma_1=0$] as well as the ground state. 
In brief, any amount
of excited state in the superposition increases the entropy.
This is intuitively expected, since it can be shown that 
the expectation of energy in the state (\ref{super1}) is give by:
$<H(\a_1,\b_1,\gamma_1)>=\f{\omega_-}{2}
\le[ 
\a_1^2(1+3R^2) + \b_1^2 ( 3+ R^2) + \gamma_1^2 (1+R^2)
\ri]$, from which it follows that   
$<H(\a_1,\b_1,\gamma_1)> - <H(0,0,1)> 
= \f{\omega_-}{2}\le( \a_1 R^2 + \b_1^2 \ri) \geq 0~$.
That is, the expectation of energy is least for the ground state; 
and higher energies are normally associated with higher entropies.   
%
%In Fig.1,
%we show the graph of our result for $\a_1=\b_1=1/\sqrt{2},
%\gamma_1=0$, which is when the excited states contribute maximum. 
%
\begin{figure}%[htb]
\label{graph1}
\begin{center}
\epsfxsize 3.00 in
\epsfysize 2.30 in
\epsfbox{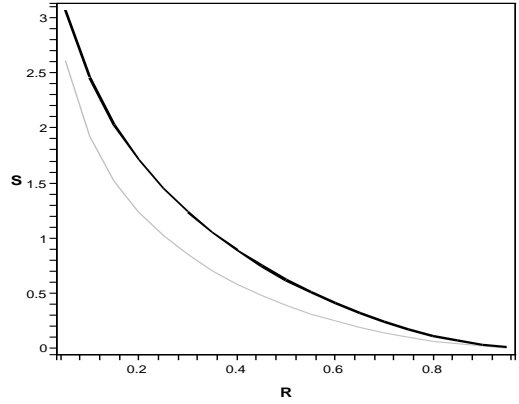}
\caption{Plots of the entanglement entropy of the excited state
$S(1/\sqrt{2},1/\sqrt{2},R)$ (black curve) and that of the ground
state $S(0,0,R)$ (grey curve) vs. $R$.  Note that 
$S(1/\sqrt{2},1/\sqrt{2},R) > S(0,0,R)$
%excited state entropy is greater than the ground state entropy 
for all $R<1$.}
\end{center}
\end{figure}
\section{Summary and Outlook}
In this paper, we have shown that the entanglement entropy of two
coupled HOs, with coordinates of one traced over, is more for excited
states compared to when they are both in their ground states.  We
would like to extend our results to $N$ oscillators, with $n<N$ of
them being traced over. This would enable one to compute the
entanglement entropy of a free scalar field in flat space-time when its
degrees of freedom inside a given region are traced over, and check
whether it is proportional to the area of the bounding surface
\cite{sdss}.  It would also be interesting to extend the results to BH
space-times with the surface mentioned above coinciding with its event
horizon. We hope to report on these and related issues in the near
future.

{\bf Note added:} After the submission of this paper to 
the {\it Canadian Journal of Physics}, we
extended the work (reported here) to the free scalar field in flat
space-time in Ref. \cite{sdss}. We have shown that the entanglement
entropy is proportional to area when the scalar field degrees of
freedom are in generic coherent states, and first excited state,
although in the latter case, the entropy increases manyfold.

%\vs{-.3cm}
\section{Acknowledgments}
SD thanks the organizers of Theory Canada-I for hospitality. We thank
R. K. Bhaduri, A. Dasgupta, J. Gegenberg, A. Ghosh, V. Husain,
G. Kunstatter, S. Nag, T. Sarkar and R. Sorkin for useful discussions
and M. Srednicki for useful correspondence. SS thanks the Physics
Dept.,%artment, 
Univ. of Lethbridge for hospitality, where part of
the work was done. This work was supported in part by 
%NSERC, Canada.
the Natural Sciences and Engineering Research Council of Canada.

%\vs{-.3cm}

\end{document}